# Extreme Inundation Statistics on a Composite Beach

Ahmed Abdalazeez [1,*], Ira Didenkulova [1,2], Denys Dutykh [3] and Céline Labart [3]

1 Department of Marine Systems, Tallinn University of Technology, Akadeemia tee 15A, Tallinn 12618, Estonia
2 Nizhny Novgorod State Technical University n.a. R.E. Alekseev, Minin str. 24, Nizhny Novgorod 603950, Russia; Irina.Didenkulova@taltech.ee
3 Univ. Grenoble Alpes, Univ. Savoie Mont Blanc, CNRS, LAMA, 73000 Chambéry, France; Denys.Dutykh@univ-smb.fr (D.D.); Celine.Labart@univ-smb.fr (C.L.)
* Correspondence: ahabda@taltech.ee.

**Abstract:** The runup of initial Gaussian narrow-banded and wide-banded wave fields and its statistical characteristics are investigated using direct numerical simulations, based on the nonlinear shallow water equations. The bathymetry consists of the section of a constant depth, which is matched with the beach of constant slope. To address different levels of nonlinearity, the time series with five different significant wave heights are considered. The selected wave parameters allow also seeing the effects of wave breaking on wave statistics. The total physical time of each simulated time-series is 1000 hours (~360000 wave periods). The statistics of calculated wave runup heights are discussed with respect to the wave nonlinearity, wave breaking and the bandwidth of the incoming wave field. The conditional Weibull distribution is suggested as a model for description of extreme runup heights and assessment of extreme inundations.

**Keywords:** wave statistics; wave runup; numerical modelling; nonlinear shallow water theory; wave breaking; freak runups

## 1. Introduction

Estimating extreme runup events in coastal zones is an important task. Flood prediction received a lot of attention in recent decades, in order to reduce hazard risks in coastal zones [1-4]. The statistical distribution of wave runup characteristics is influenced by many factors, such as topography and coastline, nonlinearity and wave breaking [5-7].

Also, some individual waves at the coast may be unexpected, extreme and hazardous. This regards sneaker waves or freak wave runups [8-10]. Such extreme events at the coast often lead to human injuries and fatalities, when people are washed off to the sea from a gentle beach or from coastal rocks or sea walls, and damage of coastal structures. During the period of 2011-2018, there were cases when freak wave runups (unrelated to tsunami) washed cars and motorcycles into the sea and damaged houses and buildings in the coastal zone [10]. These events correspond to the very tails of the statistical runup height distribution and their analysis requires extremely large datasets.

Previous studies have employed different methods to study the statistics of long wave runup, including numerical models, experiments, and field measurements.

Theoretically, [11] studied the statistical characteristics of long waves on a beach of constant slope using an analytical solution of the nonlinear shallow water theory. The study revealed that the runup height was distributed according to the Rayleigh distribution, if the incident wave elevation was described as having a normal distribution and a narrow-band spectrum. In terms of the statistical moments of the moving shoreline on a beach of constant slope, this study asserts that the kurtosis is positive for weak amplitude waves and negative for strongly nonlinear waves, whereas the skewness is always positive. Later [12] showed that for the description of even non-breaking waves the Gaussian distribution is inappropriate. Both theoretical studies had a number of assumptions, which were putting in question the applicability of these results.

Experimentally, [13] tried to reproduce theoretical results of [11] in the wave flume at Warwick University. However, they could not generate "pure" Gaussian wave field. Moreover, the generated waves were affected by capillary effects. Thus, the only result [13] could reproduce regarded an increase in the mean sea-level elevation with an increase in wave nonlinearity attributed to the known phenomenon of wave set-up. They also found that the values of the statistical moments of wave runup (skewness and kurtosis) were similar to those of the incident wave field. [14] studied statistics of narrow-band and wide-band wave runups in the Large Wave Flume of the University of Hannover, Germany. They found that wave fields with a narrow-band spectrum were associated with a higher loss of the wave energy compare to the waves with a wide-band spectrum. However, their experimental records were not long enough to discuss freak runups.

In the field measurements, [15] studied runup heights, measured on a wide spectrum of sandy beaches in New South Wales; they found that runup was distributed according to the Rayleigh distribution. [16, 17] studied wave runup at Canadian and Australian coasts and demonstrated that wave runup deviates from the Gaussian distribution. Although some of these conclusions were similar to those of [11-13], it was not possible to put direct correspondence between these works due to a number of reasons. First, the field measurement studies lacked information about an incident wave field. Second, they had a different bathymetry and coastal topography, deviating from the ideal plane beach. Third, the data included an error associated with measurement techniques.

However, the main issue, which complicates comparison of theoretical [11, 12] and experimental [13, 14] results, is the insufficient length of the experimental time-series, which do not support analysis of extreme runup statistics. Potentially, this issue can be overcome nowadays with the use of IP high-resolution cameras permanently installed on a beach and associated techniques [18-24]; however, we have not seen such works yet.

In this paper we cover the existing gap in long-term experimental records by using digital data obtained with intensive numerical computations. This approach has clear advantages. It gives control on the initial wave field offshore and allows checking the applicability of the approximated analytical results by [11, 12] to a more realistic bathymetry: plane beach merged with the flat bottom.

The paper is organized as follows. In section 2, the numerical model, based on nonlinear shallow water equations is described. The statistical moments and the distribution functions of random wave and runup fields, as well as distribution functions of wave and runup heights, are described in detail in Section 3. Then, the main results are summarized in Section 4.

## 2. Numerical Model

In this section, the 1D nonlinear shallow water model, which represents the mass and momentum conservation, is briefly described:

$$D_t + (Du)_x = 0, \tag{1}$$

$$\frac{\partial}{\partial t}(Du) + \frac{\partial}{\partial x}\left(Du + \frac{g}{2}D^2\right) = gD\frac{dh}{dx}. \tag{2}$$

Here $D = h + \eta$ is the total water depth, $\eta(x, t)$ is the water elevation, with respect to the still water level, x is the coordinate directed onshore, and $t$ is time, $h(x)$ is the unperturbed water depth, $u(x, t)$ is the depth-averaged water flow velocity, and $g$ is the gravitational acceleration. The dimensionless formulation can be obtained by choosing a typical water depth $h_0$ as the length scale (in this problem, the depth of the constant section can be taken as $h_0$ $\sqrt{gh_0}$, $\sqrt{gh_0}$ as the velocity scale and $h_0/\sqrt{gh_0}$ $h_0/\sqrt{gh_0}$ as the time scale. The dimensionless equations take the form of equations (1), (2) with $h_0 = 1$ and $g = 1$. All computations reported in this study were performed in the dimensionless formulation.

The modelling is performed in the framework of equations (1), (2), which are solved using a modern shock-capturing finite volume method. Although the shallow water model does not pursue the wave breaking and undular bore formation in a general sense (including the water surface overturning), it allows shock-wave formation and propagation with the speed given by Rankine-Hugoniot jump conditions, which, to some extent, approximates wave breaking. The numerical scheme is second order accurate, thanks to the spatial reconstruction (UNO2). For details, see [25].

In this simulation, the corresponding bathymetry (Figure 1) set-up is used: the flat part of the flume matches the beach of constant slope:

$$h(x) = \begin{cases} h_0, & x \in [a,b] \\ h_0 - (x-b)\tan\alpha, & x \in [b,c] \end{cases} \quad (3)$$

where $h_0$ is the constant water depth, kept at 3.5 m for all simulations, [a, c] are the left and right boundaries of the numerical flume, [b] is the point where the slope starts, and $\tan\alpha$ = 1:6 is tangent of the bottom slope. For simplicity, the left boundary is taken (a = 0). The length of the section of constant depth is b = 251.5 m, and the right limit of the numerical flume is taken c = 291.5 m. The number of spatial grid points along the distance between [a] and [c] is fixed and equal to 1000 for all experiments. The time step is chosen to satisfy the Courant–Friedrichs–Lewy condition for all considered significant wave heights. The spatial grid step is, therefore, 25 cm, which corresponds to 4 cm vertical resolution for runup height. This was done in order to limit simulation time, when running 10 000 hrs of physical time of wave propagation. However, this also implies that we have a low resolution and not so reliable statistics especially for small amplitude waves $H_s$ = 0.1 m. In a similar manner to the significant wave height, $H_s$, the significant runup height, $R_s$, is introduced as an average of one third of the largest runup heights in the time-series. The significant runup height for this small amplitude case is $R_s$ = 0.23 m, so even in this case the resolution is low, but considerable.

Of course, the number of extreme runups in this resolution is also somehow underrepresented, however all qualitative and comparative conclusions of this study still hold on.

*2.1. Boundary Condition*

On the left extremity x = a of the computational domain, the Dirichlet boundary condition on the total water depth component $D(a, t) = h_0 + \eta_0(t)$ of the solution (D, Du) is imposed. Namely, the free surface elevation function, $\eta_0$, is drawn from a narrow- or wide-band Gaussian signal depending on the experiment. This data turns out to be enough to obtain a well-posed initial boundary-value problem provided that the flow is subcritical at the point x = a, i.e. $|u(a,t)| < \sqrt{gD(a,t)}$ , which is always the case for Riemann waves (see [26] for the rigorous mathematical justification of this fact in case of transparent boundary conditions). The boundary conditions are implemented in the finite volume scheme according to the method described in [27], see also [28] for more details on the application to the nonlinear shallow water equations).

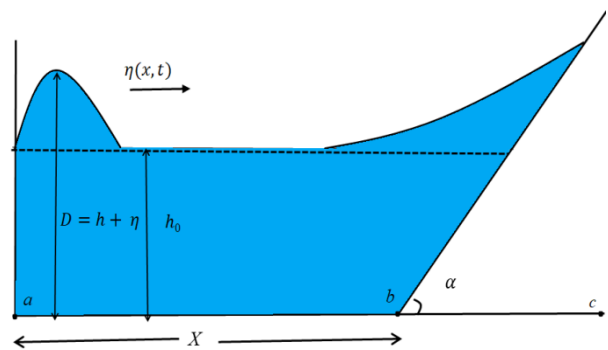

**Figure 1.** Bathymetry sketch of numerical experiment.

The considered boundary condition (wave field offshore) is distributed by the Gaussian distribution:

$$f(\xi) = \frac{1}{\sigma\sqrt{2\pi}} e^{-\frac{1}{2}\left(\frac{\xi-\mu}{\sigma}\right)^2}, \quad (4)$$

where, $\sigma$ is a standard deviation, and $\mu$ is a mean value of the distribution. To ensure this, all individual time-series have been verified by the Kolmogorov–Smirnov test [29].

The spectrum of the generated waves is

$$S(f) = \frac{S_0}{\sqrt{2\pi \Delta f / f_0}} e^{-\frac{(f/f_0 - 1)^2}{2\Delta f / f_0}}, \quad (5)$$

where $f$ is the wave frequency, $\Delta f$ is the frequency band, $f_0 = 0.1$ Hz is the central frequency, and $S_0$ is the constant, which is calculated in order to achieve the desired $H_s$.

In this work, the case with $\Delta f/f_0 = 0.1$ is referred to narrow-band spectrum, while the case with $\Delta f/f_0 = 0.4$ is attributed to the wide-band spectrum. In order to study the influence of wave nonlinearity during wave propagation to the coast, waves of different significant wave heights, which is calculated as averaged of one third of the largest wave heights in the time-series ($H_s$ = 0.1 m, 0.2 m, 0.3 m, 0.4 m, and 0.5 m) are considered. The calculated time-series for each $H_s$ is 1000 hours (360 000 wave periods). Parallel computations have facilitated the calculation of the statistics of wave runup characteristics for 5000 hours, for each bandwidth, and 10000 hours in total. The numerical computations have been carried out in MATLAB and run on a cluster containing 28 cores.

Parameter of the nonlinearity for generated waves is estimated as $H_s/h_0$ and is changing from 0.03 to 0.14. The characteristic parameter $kh_0 = 0.38$ is at the border of validity of the shallow water theory taking into account the horizontal extent of the wave tank. The phase velocity relative error committed by non-dispersive theory for $kh_0 = 0.38$ is only 2.3%. Thus, at the end of the numerical wave tank the difference between wave crest positions (between dispersive and non-dispersive models) is less than 10%. Since the focus of this paper is on wave runup, the choice of the theory is justifiable. The choice of wave parameters allows us to see the effects of wave breaking on statistics of their runups. The type of wave breaking is defined by the Iribarren number [5]:

$$Ir = \frac{\alpha}{\sqrt{H/L}}, \quad (6)$$

where $H$ is the wave height, and $L$ is the characteristic wavelength offshore. It is surging or collapsing for $Ir \geq 3.3$, plunging for $0.5 \leq Ir \leq 3.3$, and spilling for $Ir \leq 0.5$. In our dataset, only the first two types of wave breaking, surging or collapsing and plunging, are observed. For $H_s/h_0 = 0.03$, less than 1% of waves experience plunging breaking, while most of waves are surging. With an increase in $H_s/h_0$ the percentage of plunging waves increases. For $H_s/h_0 = 0.06$, 32-35% of wave are plunging, for $H_s/h_0 = 0.09$, 61-65% of wave are plunging, for $H_s/h_0 = 0.11$, 71-76% of waves are plunging, and for the most nonlinear case $H_s/h_0 = 0.14$, 85-88% of waves are plunging.

## 3. Data analysis and results

Figure 2 shows probability density functions (PDF) of narrow-band and wide-band wave fields for different nonlinearities, $H_s/h_0$. The data of the narrow-band spectra, $\Delta f/f_0 = 0.1$ are shown by triangles (different colors correspond to different nonlinearities), while the corresponding Gaussian distribution ($\mu = 0$, $\sigma = 0.25$) is shown by the black solid line. The data of the wide-band spectra, $\Delta f/f_0 = 0.4$ are shown by pluses, and the corresponding Gaussian distribution ($\mu = 0$, $\sigma = 0.27$) is shown by the red solid line. It can be seen that the generated waves are well described by the Gaussian distribution, which has zero mean, skewness and kurtosis for all nonlinearities, $H_s/h_0$.

To describe the wave statistics in Figure 2, the Rayleigh distribution, which is well used for this type of problem [5], is applied:

$$f(\xi) = \begin{cases} \frac{\xi}{\lambda^2} e^{-\xi^2/(2\lambda^2)}, & \xi \geq 0 \\ 0, & \xi < 0 \end{cases}, \quad (7)$$

where $\xi$ is a data vector, $\lambda$ is the scale parameter. For a better fit, a two-parameter Weibull distribution is also considered:

$$f(\xi) = \begin{cases} \frac{k}{\lambda}\left(\frac{\xi}{\lambda}\right)^{k-1} e^{-(\xi/\lambda)^k}, & \xi \geq 0 \\ 0, & \xi < 0 \end{cases}, \quad (8)$$

where $\lambda$ is the scale parameter, and $k$ is the shape parameter.

The wave height distributions of both narrow-band and wide-band wave fields are shown in Figure 5. As expected, the narrow-band data are well described by a Rayleigh distribution ($\lambda = 0.5$), although a Weibull distribution gives a slightly better fit ($\lambda = 0.74$, $k = 2.27$). The data of wide-band spectra tend to be distributed according to a Weibull distribution ($\lambda = 0.71$, $k = 2.06$).

The waves which are twice higher than the significant wave height ($H/H_s \geq 2$) are the so-called freak waves. It can be seen from Figure 3 that the probability of the freak wave occurrence in the initial wave field is higher for narrow-band signals than for wide-band ones.

The calculated significant runup heights $R_s$ for narrow-band and wide-band signals are shown in Figure 4. Interesting to see that $R_s$ for wide-banded waves is always higher than for narrow-banded waves, which can be explained by higher variability in wave periods for wide-banded waves. Also, figure 4 indirectly shows us how many of our waves are breaking. The wave runup height, at which the first wave breaking occurs in the wave trough can be estimated as $R_{cr}/h_0 = g(\alpha T/(2\pi))^2/h_0 = 0.2$, see for details [30]. This means that our case of "small" nonlinearity $H_s/h_0 = 0.03$ is very little affected by wave breaking (< 1% according to Iribarren criterion). The case of $H_s/h_0 = 0.06$ is affected by wave breaking only for extreme runups (32-35% according to Iribarren criterion). In the case of $H_s/h_0 = 0.09$, more than a half of waves are breaking (61-65% % according to Iribarren criterion). However, in the cases of $H_s/h_0 = 0.11$ and $H_s/h_0 = 0.14$ the majority of waves are breaking.

Figure 5 shows the probability distribution functions of runup oscillations, $r/R_s$ for initial Gaussian narrow-banded and wide-banded wave signals. It can be seen from Figure 5a, that runup oscillations of narrow-banded waves are no longer distributed by a normal distribution, and are slightly shifted to the right towards larger positive values with an increase in nonlinearity. Partially this effect was observed both theoretically for an infinite plane beach [11, 12] and experimentally [13, 14]. What is interesting and peculiar is a strong deformation of the distribution itself. In addition, the tails of these distributions are much thinner than of Gauss, and reflect a relatively weak probability of extreme floods for narrow-banded waves.

The distributions of runup oscillations of initial wide-band signal are also shifted to the right towards higher runups with an increase in nonlinearity, but this shift is much larger compared to the one of the narrow-band signal. Moreover, the tails of these distributions are much thicker than those for narrow-band data, and are rather close to the normal distribution, which corresponds to a relatively large probability of extreme floods for wide-banded waves.

It can also be seen that for both narrow-banded and wide-banded waves, the probability of large waves decreases with an increase in wave nonlinearity, which can be explained by wave breaking.

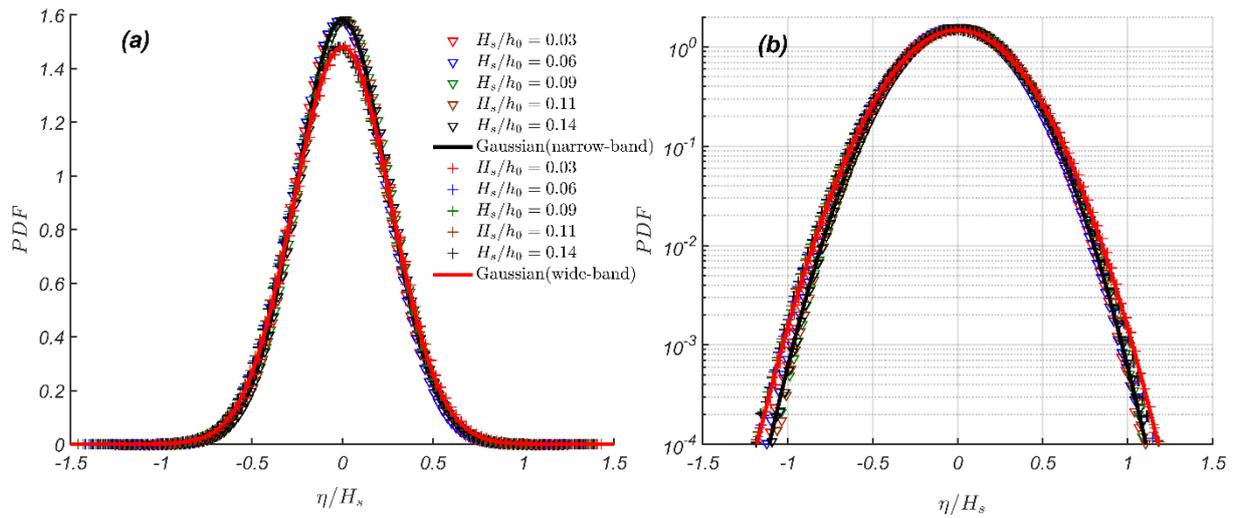

**Figure 2.** Probability density functions of normalized narrow-band and wide-band wave fields offshore for different nonlinearities, $H_s/h_0$ in linear (left) and logarithmic (right) scales. Solid lines correspond to Gaussian distributions fitted to the corresponding datasets, shown with a red color for wide-band data and with black color for narrow band data.

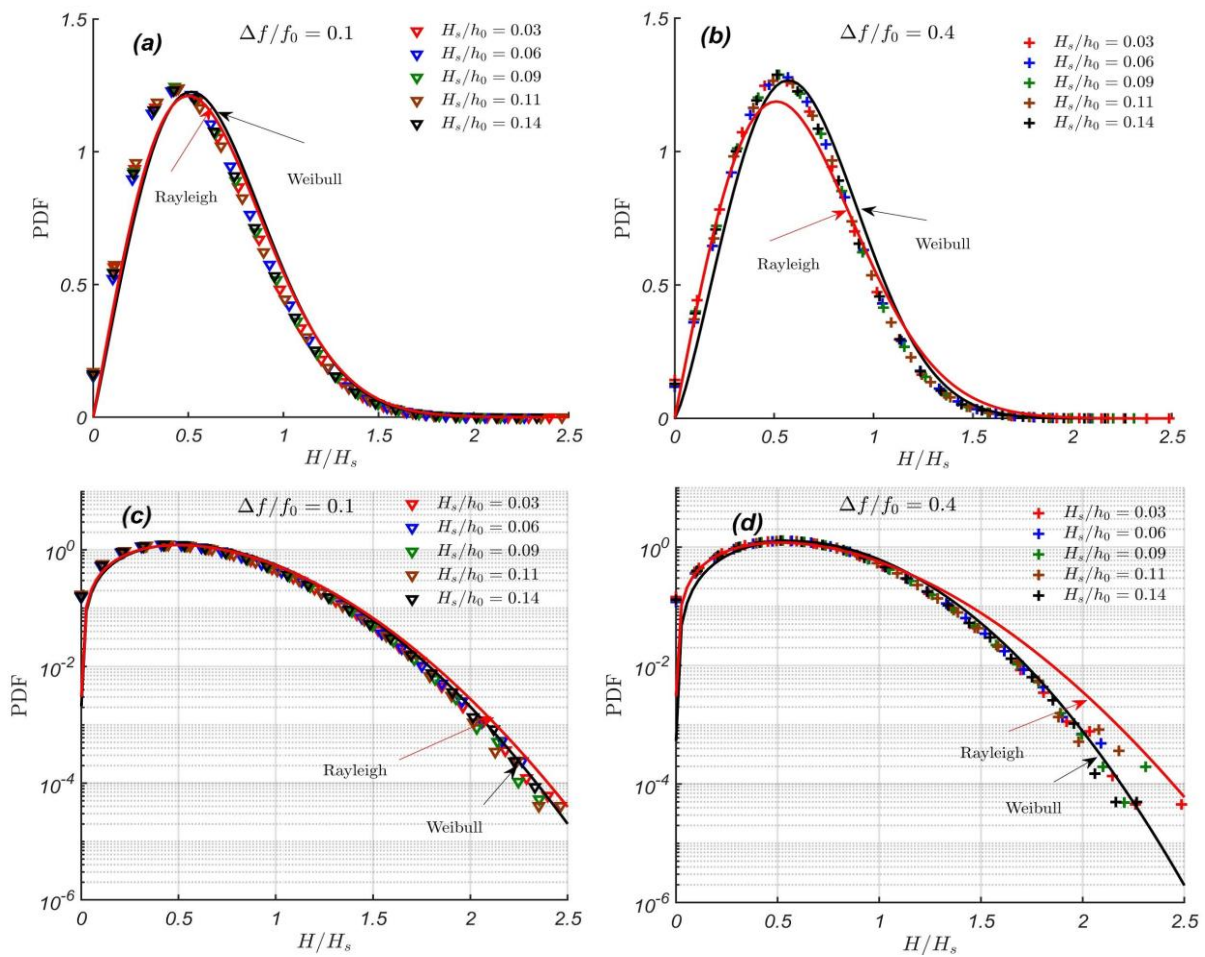

**Figure 3.** Probability density functions of normalized trough-to-crest wave heights of the initial narrow-band (a) and (c), and wide-band (b) and (d) wave fields for different nonlinearities, $H_s/h_0$ in linear (top) and logarithmic (bottom) scales. Red solid line corresponds to the Rayleigh distribution; black solid line corresponds to the Weibull distribution fitted to the corresponding dataset.

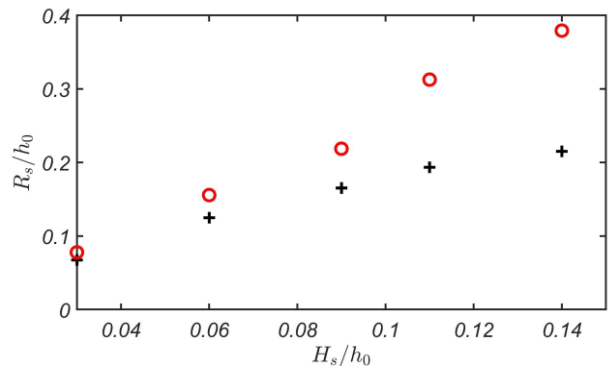

**Figure 4.** Significant runup height, $R_s$ for wide-band (red circles) and narrow-band (black crosses) signals for different nonlinearities.

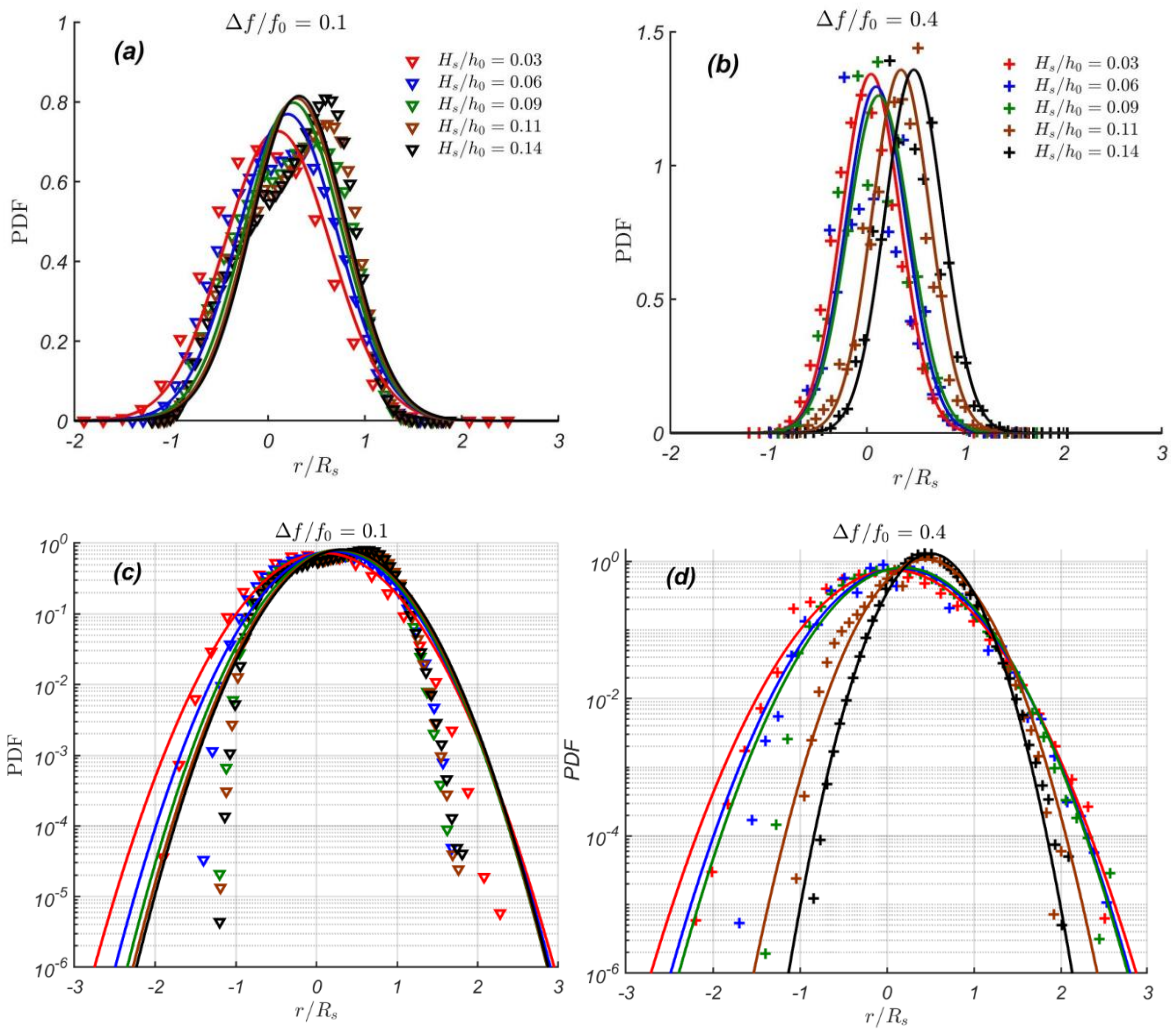

**Figure 5.** Probability density functions of runup oscillations, normalized by a significant runup height, $R_s$, for different nonlinearities for narrow-banded (a) and (c), and wide-banded (b) and (d) waves in linear (top) and logarithmic (bottom) scales. Solid lines correspond to Gaussian distributions, fitted to the corresponding datasets, using the matching colors.

These effects can also be seen in Figure 6, which shows the statistical moments of narrow-banded and wide-banded waves offshore, normalized by $H_s$, and the corresponding runup oscillations on a beach, normalized by $R_s$. The statistical moments, mean, variance, skewness, and (normalized) kurtosis are calculated as:

$$<\xi> = \frac{1}{n}\sum_{i=1}^{n} \xi_i, \qquad \sigma^2(\xi) = \frac{1}{n}\sum_{i=1}^{n}(\xi_i - <\xi>)^2, \qquad (9)$$

$$Sk(\xi) = \sum_{i=1}^{n}\frac{1}{n\sigma^3(\xi)}(\xi_i - <\xi>)^3, \qquad Kurt(\xi) = \sum_{i=1}^{n}\frac{1}{n\sigma^4(\xi)}(\xi_i - <\xi>)^4 - 3, \qquad (10)$$

where $\xi$ is a data vector, and $n$ is its length.

Noteworthy, the mean, skewness and kurtosis of both narrow-banded and wide-banded wave fields are zero, which provide the desired Gaussian statistics. Regarding runup oscillations, one can see that for both narrow- and wide-banded waves the mean of runup oscillations rises with the nonlinearity, which reflects the known effect of wave set-up on a beach. For small-amplitude waves, the set-up for narrow-banded waves is larger than for wide-banded ones, while for large amplitude waves, affected by wave breaking, it is the opposite. For wide-banded waves, the variance decreases with an increase in nonlinearity, while for narrow-banded waves it changes non-monotonically. The higher moments, skewness and kurtosis of runup oscillations for waves with narrow-band spectrum are negative, while for waves with wide-band spectrum they are sign-variable. Also for the narrow-banded waves, the skewness decreases with an increase in wave nonlinearity, while kurtosis changes non-monotonically with an increase in wave nonlinearity. Moreover, for wide-banded waves, both skewness and kurtosis change non-monotonically with an increase in nonlinearity. This somehow only partially corresponds to the theoretical findings in [11], where the kurtosis was positive for weak amplitude waves and negative for strongly nonlinear waves, while the skewness was always positive. However, in the experimental study of [13], the skewness was both positive and negative. It is also important to say that for all four moments one can see different dynamics for small-amplitude non-breaking or almost non-breaking waves and large-amplitude waves, strongly affected by wave breaking.

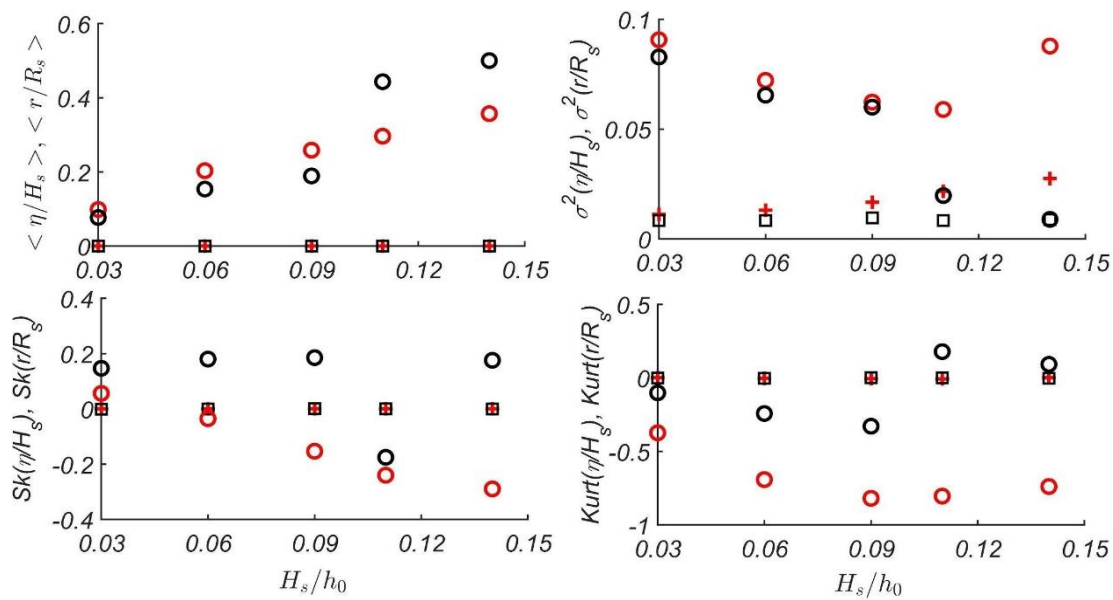

**Figure 6.** Statistical moments of runup oscillations (normalized by $R_s$) of narrow-banded (red circles) and wide-banded (black circles) waves on a beach, $r$, versus nonlinearity, $H_s/h_0$. Statistical moments of narrow-band and wide-band wave fields offshore (normalized by $H_s$) are shown by red crosses and black squares respectively.

Runup oscillations deviate from the Gaussian distribution even for weak-amplitude waves (see Figure 6). With an increase in nonlinearity, all statistical moments of runup oscillations change. It can be seen that statistical moments of narrow-banded irregular waves (except kurtosis) change with $H_s$ monotonically, while for the wide-banded waves, they vary non-monotonically (except mean values).

The large (extreme) wave runup heights, $R_{extrm} = R/R_s \geq s$, where s is some threshold value, somehow behave similar to a conditional Weibull law whose density is given by Eq. (11):

$$f(R_{extrm}) = \begin{cases} \dfrac{k}{\lambda}\left(\dfrac{R_{extrm}}{\lambda}\right)^{k-1} e^{-(R_{extrm}/\lambda)^k + (s/\lambda)^k}, & R_{extrm} \geq s \\ 0, & R_{extrm} < s \end{cases} \tag{11}$$

A conditional Weibull law is characterized by three parameters: the shape k, the scale $\lambda$ and the threshold s. Given the data $(R_{i\,extrm})$ = 1…n, s is fixed and k and $\lambda$ are computed by maximum likelihood estimator. The scale parameter, $\lambda$ can be obtained from Eq. (12):

$$\lambda = \left(\frac{1}{n}\sum_{i=1}^{n}(R_{i\,extrm}^k - s^k)\right)^{\frac{1}{k}}, \tag{12}$$

where *n* is a number of extreme wave runups. In order to obtain the shape parameter, *k*, one should solve Eq. (13):

$$\frac{1}{k}(I_n - s^k) + (\ln s)s^k + M_n(I_n - s^k) - V_n = 0, \tag{13}$$

$$M_n = \frac{1}{n}\sum_{i=1}^{n}\ln R_{i\,extrm} \tag{14}$$

$$I_n = \frac{1}{n}\sum_{i=1}^{n}R_{i\,extrm}^k, \tag{15}$$

$$V_n = \frac{1}{n}\sum_{i=1}^{n}(\ln R_{i\,extrm})R_{i\,extrm}^k \tag{16}$$

Similarly to freak waves, the waves on a beach, whose runup height is twice larger than the significant runup height ($R/R_s \geq 2$), we call freak runups. On gentle beaches, such freak runups are manifested as sudden floods and may result in human injuries and fatalities [8-10].

Figure 7 shows probability distribution functions of large runup heights ($R \geq 0.7\ R_s$), for narrow-band and wide-band spectra, for different nonlinearities. It can be seen in Figure 7, the tails of distributions for runup heights corresponding to freak events for narrow-banded waves decay much faster than those for wave heights offshore (except waves of weak amplitude with $H_s/h_0$ = 0.03), which means that for narrow-banded waves the probability of freak runup occurrence on a beach is less than the probability of freak wave occurrence in the sea coastal zone and a gentle beach works as some kind of "filter" for narrow-banded freak events. This is also manifested in the numbers of actual freak events, given in Table 1. It can be seen that for non-breaking waves of the smallest amplitude $H_s/h_0$ = 0.03, the number of freak events on a beach was reduced twice compared to the original number of freak waves offshore, while for waves of larger amplitude, which were affected by the wave breaking, there were no freak runups at all.

In contrast, for wide-banded waves the probability of freak events on a beach is more or less the same as in the sea coastal zone and may even be higher (Figure 7). The number of freak runups for small non-breaking wide-banded waves increased twice as compared to the original number of freak waves offshore (see Table 1). With an increase in wave amplitude (and consequently, wave breaking), the number of freak runups on a beach decreases, however for waves of moderate amplitude, the number of freak runups is still larger than the number of freak waves offshore, while for waves strongly affected by the wave breaking ($H_s/h_0$ = 0.11 and 0.14), the number of freak runups on a beach suddenly drops down (see Table 1).

**Table 1.** The number of freak events in the sea coastal zone and on a beach for different wave regimes.

|  | *Δf/f₀* = 0.1 | | | *Δf/f₀* = 0.4 | | |
|---|---|---|---|---|---|---|
| $H_s/h_0$ | Number of waves | Freak waves offshore | Freak runups | Number of waves | Freak waves offshore | Freak runups |
| 0.03 | 362255 | 125 | 61 | 389232 | 51 | 118 |
| 0.06 | 362380 | 117 | 0 | 389385 | 45 | 76 |
| 0.09 | 362096 | 89 | 0 | 389444 | 49 | 62 |
| 0.11 | 362319 | 88 | 0 | 389263 | 53 | 2 |
| 0.14 | 362302 | 102 | 0 | 389728 | 34 | 1 |

Probability of extreme wave runups on a beach is noticeably higher for waves with wide-band spectrum, than for waves with narrow-band spectrum (see Figure 7), although the probability of extremes wave heights in the wave field offshore is significantly higher for narrow-banded waves than for wide-banded ones (see Figure 3, Table 1).

The probability of extreme runup formation changes with the wave nonlinearity. It is decreasing with an increase in wave nonlinearity for wide-banded waves and changes non-monotonically with nonlinearity for narrow-banded waves. It is also interesting to see that the tails of distributions in Figure 9 are somehow gathered into clusters and can be separated in two groups for "relatively large $H_s$" and "relatively small $H_s$", where the "small $H_s$" group is always higher than the "large $H_s$" group. The latter holds for both narrow-banded and wide-banded waves and can be explained by the wave breaking.

The corresponding data of wave runup heights are also approximated by a conditional Weibull distribution [Eq. (11)], which gives reasonable results and can be used to evaluate the probability of freak runups. Here the threshold s is selected as 0.7 and the calculated parameters *k* and *λ* are given in Table 2.

**Table 2.** Parameters of conditional Weibull distribution fitted to the corresponding datasets in Figure 7.

|  | *Δf/f₀* = 0.1 | | *Δf/f₀* = 0.4 | |
|---|---|---|---|---|
| $H_s/h_0$ | k | λ | k | λ |
| 0.03 | 2.747 | 0.886 | 0.76 | 0.116 |
| 0.06 | 3.6 | 0.92 | 1.43 | 0.48 |
| 0.09 | 4.06 | 0.89 | 2.58 | 0.86 |
| 0.11 | 3.08 | 0.777 | 2.6 | 0.772 |
| 0.14 | 3.08 | 0.72 | 2.718 | 0.762 |

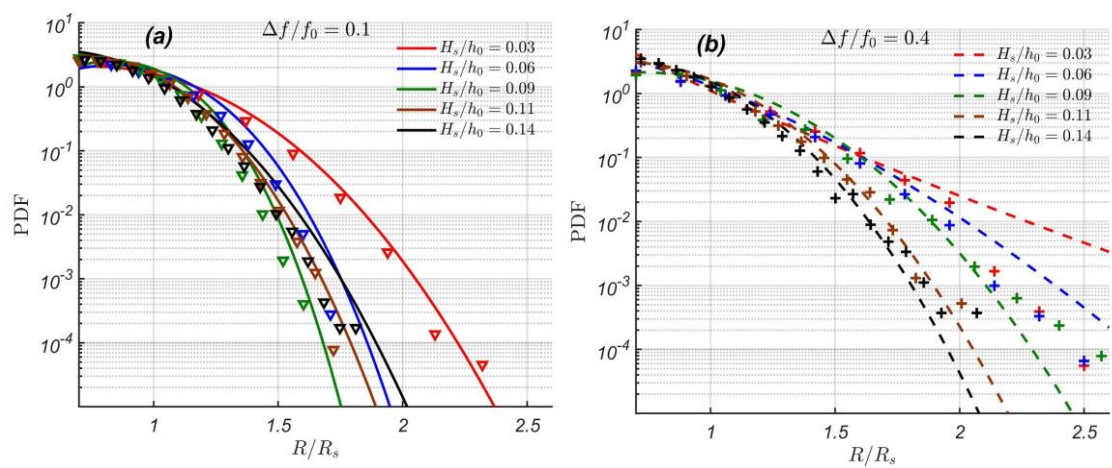

**Figure 7.** Probability density functions of large runup heights ($R \geq 0.7R_s$) for (a) narrow-banded (triangles) and (b) wide-banded (pluses) waves. Lines correspond to conditional Weibull distributions [Eq. (11)], fitted to the narrow-band (solid lines) and wide-band (dashed lines) datasets, using the matching colors.

## 4. Conclusions

In this paper, irregular waves runup on a plane beach is studied by means of direct numerical simulations. The numerical model is based on the nonlinear shallow water equations and is of the second order of accuracy. The corresponding bathymetry consists of the section of constant depth, which is matched with the beach of a constant slope. The irregular waves are represented by the Gaussian wave field with spectra of two different bandwidths, which are referred to as narrow-banded and wide-banded waves. To address different levels of wave nonlinearity, time-series with five different significant wave heights are considered. The selected wave regimes represent (i) non-breaking waves, (ii) waves slightly affected by wave breaking, (iii) moderate wave breaking and (iv) significant wave breaking, when the majority of waves are breaking. Each of these time-series has a duration of 1000 hours (360 000 wave periods).

The heights of narrow-banded waves are well described by Rayleigh distribution, while heights of wide-banded waves are described by Weibull distribution irrespective of the wave nonlinearity. However, for wide-banded waves the very tails of these distributions show larger variability than for narrow-banded ones.

As expected, the runup oscillations are not Gaussian, which confirms results of many previous studies, both theoretical [11, 12] and experimental [13, 16, 17]. For both narrow-band and wide-band cases, one can observe the effect of wave set-up (increase in the mean value of runup oscillations), which increases with an increase in wave nonlinearity. However, for wide-banded waves this increase is significantly stronger than for narrow-banded ones.

What regards extreme, so-called "freak events", their statistics in the initial narrow-banded wave signal offshore is more representative, than on the beach ("freak runups") even for non-breaking waves. Therefore, for narrow-banded waves, gentle beaches reduce the number of freak events as compared to the sea coastal zone, and work as a 'low-pass filter' for extreme wave heights. This may explain why freak events on a beach are so unexpected [8-10]. However, for wide-banded waves, such an effect has not been observed and the probability of freak events on a beach was similar to or even larger than the one in the sea coastal zone.

The number of freak events in wide-band and narrow-band cases varies, so that increase in the bandwidth leads to a substantial increase in the number of freak events. This can be explained by higher variability in wave periods for wide-banded waves, and wave runup height is rather sensitive to these variations. In addition, the number of freak waves decreases with an increase in wave amplitude and consequently, wave breaking. The largest number of freak waves was observed for non-breaking wide-banded waves, which almost doubled the number of freak waves in the boundary condition wave record.

Finally, to describe statistics of extreme wave runup heights on a gentle beach, a conditional Weibull distribution is suggested. It gives reasonable results and may be used for assessment of extreme inundations on a beach (freak runups). In addition, in future applications the statistical analysis hereby provided might also be useful in the study of the wave run-up phenomenon in other applications, e.g. in structures placed in shallow water conditions [31, 32].

The limitation imposed by the resolution of the numerical simulations should also be taken into account. Although the number of freak waves on the beach may be somehow reduced by a coarse model resolution, the qualitative and comparative conclusion of this study should not be affected. This point will be improved in our future studies.

**Author Contributions:** Conceptualization, I.D.; methodology, I.D., D.D., C.L; software, D.D.; resources, I.D., D.D.; data curation, A.A.; writing—original draft preparation, A.A; writing—review and editing, A.A, I.D., D.D. and C.L.; numerical simulations and data analysis, A.A; Conditional Weibull distribution conceptualization and fitting C.L. All authors have read and agreed to the published version of the manuscript.

**Funding:** The study of distribution functions and the statistical moments of irregular wave runup was performed with the support of an RSF grant (16-17-00041). The numerical simulations of wave runup were supported by an ETAG grant (PUT1378). The authors also thank the PHC PARROT project (No. 37456YM), which funded the visits to France and Estonia, thus facilitating this collaboration.

**Acknowledgments:** The computations were performed using the computer cluster of the Department of Marine Systems of Tallinn University of Technology.

**Conflicts of Interest:** No conflicts.